\documentclass[10pt, conference, compsocconf]{IEEEtran}

\usepackage{mathptm}
\usepackage[utf8]{inputenc}
\usepackage[T1]{fontenc}
\usepackage[english]{babel}
\usepackage{textcomp}
\usepackage{type1cm}
\usepackage{amsmath}
\usepackage{amsfonts}
\usepackage{subcaption}
\usepackage{booktabs}
\usepackage{listings}
\usepackage{graphicx}
\let\labelindent\relax
\usepackage{enumitem}
\usepackage{color}
\usepackage{multirow}
\usepackage[symbol]{footmisc}
\usepackage{refcount}
\usepackage{scrextend}
\usepackage{float}

\usepackage[style=ieee]{biblatex}
\AtEveryBibitem{%
   \clearfield{month}%
   \clearfield{series}%
   \clearname{editor}%
   \clearlist{publisher}%
   \clearlist{location}
   \clearfield{issn}%
   \clearfield{isbn}%
   \clearfield{eventdate}%
   \clearfield{pages}%
   \clearfield{number}%
   \clearfield{volume}%
   \clearlist{language}%
   \clearfield{origlanguage}%
   \clearlist{extra}%
   \clearfield{note}
   \ifnameundef{author}
    {}
    {
    \clearfield{url}%
    \clearfield{urldate}%
    \clearfield{urlyear}%
    \clearfield{urlmonth}%
    \clearfield{doi}%
    }%
}

\usepackage{flushend}

\begin{document}

\title{Towards a Benchmarking Suite for Kernel Tuners}
\author{
    \IEEEauthorblockN{
        Jacob O. T\o rring\IEEEauthorrefmark{1}, Ben van Werkhoven\IEEEauthorrefmark{2}, Filip Petrovič\IEEEauthorrefmark{3}, Floris-Jan Willemsen\IEEEauthorrefmark{2}, Jiří Filipovič\IEEEauthorrefmark{3} and Anne C. Elster\IEEEauthorrefmark{1}
    }
    \IEEEauthorblockA{\IEEEauthorrefmark{1} Norwegian University of Science and Technology (NTNU), Trondheim, Norway\\
    }
    \IEEEauthorblockA{\IEEEauthorrefmark{2} Netherlands eScience Center, Amsterdam, Netherlands\\
    }
    \IEEEauthorblockA{\IEEEauthorrefmark{3} Masaryk University, Brno, Czech Republic\\
    }
    \{jacob.torring, elster\}@ntnu.no, \{b.vanwerkhoven, f.j.willemsen\}@esciencecenter.nl, \{fillo, fila\}@mail.muni.cz
}

\maketitle

\thispagestyle{plain}
\pagestyle{plain}

\begin{abstract}

As computing system become more complex combining CPUs and GPUs, it is becoming harder and harder for programmers to keep their codes optimized as the hardware gets updated. Autotuners try to alleviate this by hiding as many architecture-based optimization details as possible from the end-user, so that the code can be used efficiently across different generations of systems. Several autotuning frameworks have emerged, but a comparative analysis between
these related works is scarce, owing to the significant manual
effort required to port a tunable kernel from one tuner
another. 

In this article we introduce a new benchmark suite for evaluating the performance of 
optimization algorithms used by modern autotuners targeting GPUs. The suite 
contains tunable GPU kernels that are representative of real-world applications, allowing for comparisons between optimization algorithms and the examination of code optimization, search space difficulty, and performance portability. Our framework facilitates easy integration of new autotuners and benchmarks by defining a shared problem interface.

Our benchmark suite is evaluated based on five characteristics: 
convergence rate, local minima centrality, optimal speedup, 
Permutation Feature Importance (PFI), and performance portability. The results show that optimization parameters greatly impact performance and the need for global optimization. The importance of each parameter is consistent across GPU architectures, however, the specific values need to be optimized for each architecture.


Our portability study highlights the crucial importance of autotuning each application for a specific target architecture. The results reveal that simply transferring the optimal configuration from one architecture to another can result in a performance ranging from 58.5\% to 99.9\% of the optimal performance, depending on the GPU architecture. This highlights the importance of autotuning in modern computing systems and the value of our benchmark suite in facilitating the study of optimization algorithms and their effectiveness in achieving optimal performance for specific target architectures.

\end{abstract}


%
\IEEEpeerreviewmaketitle

\section{Introduction}\label{chap:introduction}

As computers have become more advanced in recent decades, there has been a significant increase in their complexity. Central Processing Units (CPUs) still form the core of modern computers, but we have seen a growing use of accelerators like Graphics Processing Units (GPUs) and co-processors to improve efficiency and performance. These accelerators can be highly effective, but they can also make the task of optimizing programs for performance increasingly difficult.

New data centers and many of the world's top supercomputers, such as the Top500 systems, have increasingly relied on High-Performance Computing (HPC) systems that use a combination of CPUs and GPUs~\cite{heldens2020landscape}. These heterogeneous systems can make code optimization a challenging task, as the architectures of the different components change rapidly.

To make programs run efficiently on these systems, programmers in 
computer science and related fields spend considerable efforts in optimization~\cite{hijma2022}. However, as the complexity of these systems grows, the task of understanding how all the system components interact becomes increasingly complex. To address this, analytical methods such as compiler optimization are used to understand the systems and turn that knowledge into rules for optimizing programs. These rules can modify large chunks of code without changing the semantics of the program, but they can also contain heuristics that may not always lead to the optimal solution. 

As architectures are constantly evolving and different, a heuristic-based rule cannot always generalize well to the entire system. In these cases, empirical methods which search for the optimal solution through trial and error can be used, this is called \textit{autotuning}. 

One of the main consideration when developing such solutions is to hide as many architecture-based optimization details as possible from the end-user, so that the code can be used efficiently across different generations of systems. The goal is to provide easy-to-use libraries and APIs that enable developers to write code that runs well on different systems without having to understand the intricacies of each system's architecture.

In recent years, several studies have been conducted that present advancements in optimization algorithms for autotuning~\cite{schoonhoven2022benchmarking}. Despite this, comparative analysis between related works is scarce, owing to the significant manual effort required to porting a tunable kernel from one tuner another. In order to effectively study the performance of optimization algorithms for autotuning, as well as to facilitate comparisons between optimization algorithms implemented in different tuners, it is necessary to have a benchmark suite that is compatible with all of these tuners. 

Existing benchmark suites, such as 
Rodinia~\cite{che_rodinia_2009}, SHOC~\cite{danalis_scalable_2010}, PolyBenchGPU~\cite{grauer-gray_auto-tuning_2012}, 
are not suitable for this purpose as they are not tunable. They typically have hardcoded thread block dimensions, but even worse, the code is often written with assumptions that the block dimensions, parallelizations, and amount of work per thread will never change. Furthermore, in many cases, the hardcoded numbers are directly dependent on the input problems, which are often unrealistically small for modern GPUs. As a result, tuning these codes requires extensive modifications.

Earlier attempts at creating tunable benchmark suites~\cite{petrovic_benchmark_2019, sund_bat_2021} are limited
regarding studying the effectiveness of optimization algorithms. The benchmark suite from Petrovi\v{c} et al.~\cite{petrovic_benchmark_2019} only supports one tuning framework, while the optimization parameters in Sund et al.~\cite{sund_bat_2021} have limited performance impact.

The "BAT 2.0" benchmark suite aims to stimulate autotuning research by offering a set of tunable kernels that are representative of those used in various real-world applications, such as machine learning, image processing, astrophysics, thermal modeling, microscopy, geospatial information systems, and radio astronomy.

This benchmark suite facilitates for comparisons between optimization algorithms from different autotuners by providing a standardized problem interface for both the autotuners and benchmarks. The benchmark suite provides general configuration space and kernel handler classes providing for easy integration towards Optuna~\cite{akiba_optuna_2019}, SMAC3~\cite{hutter_sequential_2011}, KernelTuner~\cite{van_werkhoven_kernel_2019}, KTT\cite{petrovic_benchmark_2019} as well as our own basic reference tuner. This enables the study of code optimization effectiveness, search space difficulty, performance portability, and more. With the creation of this benchmark suite, researchers can now investigate key questions about code optimization, search space, and performance portability.

\section{Background} \label{chap:background}
GPUs were originally designed for fast processing of graphics, but they have also been found to be effective accelerators for 
parallelizable AI and computational science workloads. 
After their programming environments facilitated using them as General Purpose GPUs (GPGPUs)
to speed up a variety of scientific computations, they are now a key component in many of the world's largest computing clusters. The 
optimization workload running on GPU architectures is thus 
important for a range of fields beyond video game graphics.

\subsection{Performance Portability}
Relative performance\footnote{performance of a code compared to peak performance of hardware} can vary greatly across different hardware platforms, even if a program functionally behaves the same on all of them. A problem configuration that runs well on one GPU may perform poorly on another GPU. When writing high-performance software, it is essential to utilize the hardware as efficiently as possible. One solution to this is to use libraries like ATLAS~\cite{clint_whaley_automated_2001}, which automatically tunes its configuration based on the executing hardware. This can avoid the manual effort of finding the optimal configuration for every hardware platform. FFTW~\cite{frigo_fftw_1998} also used a similar self-tuning approach to optimize its fast Fourier transforms. 

We can measure how portable these configurations are by finding the optimal configurations for our target platforms and then examine the relative performance of these optimal configurations on other systems. We can thus analyze how sensitive the configurations are to platform changes and how large the relative performance differences are between different architectures.

\subsection{Analysing characteristics of benchmarks}
\subsubsection{Feature importance}
Feature importance in ML is a technique to identify influential features of a dataset on a model's outcome~\cite{zien_feature_2009}. Different methods compute feature importance, e.g. permutation importance, feature importances from tree-based models, LASSO, etc. These methods quantify feature importance to understand a dataset's characteristics, identify redundant/irrelevant features, and guide feature selection. The feature importance scores express the dataset's characteristics and the contribution of each feature to the model's performance.

In this study, we use Permutation Feature Importance (PFI) to evaluate feature importance. PFI measures a model's performance decrease when a feature's values are shuffled to understand the feature's importance. We calculate PFI by training a Catboost~\cite{prokhorenkova_catboost_2018} Regression model on the original dataset, shuffling each feature's values, retraining the model, and comparing the original and shuffled dataset's performance metric. The PFI score for each feature is the difference between the two. PFI helps identify important features for a model's performance and can detect multicollinearity among features to prevent overfitting and unstable models.

\subsubsection{Proportion of Centrality metric}
The proportion of centrality metric introduced by Schoonhoven et al.~\cite{schoonhoven2022benchmarking} is a way to quantify the difficulty of GPU tuning. It is based on the concept of the fitness flow graph (FFG), which contains all points in the search space and creates a directed edge to a neighboring point if the neighbor has lower fitness. This means that a random walk across the FFG mimics the behavior of a randomized first-improvement local search algorithm. The expected proportion of arrivals of each minimum then gives a metric for weighting reachability of each minimum. The likelihood of arrival per local minima is computed using the PageRank node centrality, which was originally used to determine the relevance of a webpage. The PageRank values are the values of the dominant right eigenvector of the adjacency matrix of a directed graph G, rescaled such that each column adds up to 1. The metric is a measure of difficulty, it considers how likely a certain subset of "suitably good" local minima are to be visited by a local search algorithm relative to the rest. This subset is defined by the optimal fitness and the proportion p, taking the set of nodes consisting of local minima with fitness less than $(1 + p)f_{opt}$ for minimization problems, otherwise $(1 - p)f_{opt}$.

\section{Related work}\label{chap:related-work}

In the field of autotuning, the most relevant prior work includes the benchmark suite developed by Petrovi\v{c} et al.~\cite{petrovic_benchmark_2019}, Polybench-GPU~\cite{grauer-gray_auto-tuning_2012}, and Sund et al.~\cite{sund_bat_2021}. 
However, these benchmark suites have issues that limit their usefulness. The benchmark in PolyBench-GPU has small search spaces, ranging from 116 to 725 different possible configurations, which would fall below our threshold for an interesting autotuning study as many real-world applications have much larger search spaces. The benchmark suite from Petrovi\v{c} et al. only supports a single autotuning framework, while the optimization parameters in the benchmark suite by Sund et al. have limited performance impact. To address this issue, we have selected benchmarks with larger and more interesting search spaces for this new version of BAT, and also added several new benchmarks that meet our criteria. These benchmarks all give significant speedups, with the performance of the optimal configurations significantly varying between different systems.

For other related works, the Collective Knowledge framework (CK) developed by Fursin et al.~\cite{fursin_collective_2021} offers a more generalized approach to benchmarking and reproducibility. In the area of Hyperparameter Optimization (HPO) several benchmark suites have been developed, such as the high-dimensional HPO benchmark suite~\cite{wang_fedhpo-b_2022}, LassoBench~\cite{sehic_lassobench_2022}, and HPOBench~\cite{eggensperger_hpobench_2022}. 

\section{Benchmarks}

\subsection{GEMM}

Generalized dense matrix-matrix multiplication (GEMM) is part of the BLAS linear algebra 
specification, and is one of the most widely-used GPU kernels. 
The GEMM kernel included in BAT is from CLBlast~\cite{clblast}, a tunable OpenCL BLAS library.
GEMM implements the multiplication of two matrices, $A$ and $B$:
\begin{equation}\nonumber
C = \alpha A \cdot B + \beta C
\end{equation}
where $\alpha$ and $\beta$ are scalars and $C$ is the output matrix. 
The CLBlast GEMM kernel is tunable with the parameters shown in Table~\ref{tab:gemm-parameters}.  \verb|MWG| and \verb|NWG| control the amount of work assigned to each thread block. \verb|MDIMC| and  \verb|NDIMC| describe the size of thread block, while \verb|MDIMA| and \verb|MDIMB| control shared memory usage, \verb|VWM| and \verb|VWN| are the vector widths used for loading from and storing to global memory, and \verb|SA| and \verb|SB| enables or disables the use of shared memory for elements in $A$ and $B$.

\begin{table}[ht]
    \caption{Tunable parameters -- GEMM kernel in BAT.}
    \centering
    \begin{tabular}{l|l|l}
    \toprule
    Parameter & Values & \# \\
    \midrule
 \verb|MWG|  & $\{$16, 32, 64, 128$\}$ & 4 \\
 \verb|NWG|  & $\{$16, 32, 64, 128$\}$ & 4 \\
 \verb|MDIMC|  & $\{$8, 16, 32$\}$ & 3 \\
 \verb|NDIMC|  & $\{$8, 16, 32$\}$ & 3 \\
 \verb|MDIMA|  & $\{$8, 16, 32$\}$ & 3 \\
 \verb|NDIMB|  & $\{$8, 16, 32$\}$ & 3 \\
 \verb|VWM|  & $\{$1, 2, 4, 8$\}$ & 4 \\
 \verb|VWN|  & $\{$1, 2, 4, 8$\}$ & 4 \\
 \verb|SA|  & $\{$0, 1$\}$ & 2 \\
 \verb|SB|  & $\{$0, 1$\}$ & 2 \\
 \bottomrule
    \end{tabular}
    \label{tab:gemm-parameters}
\end{table}

\subsection{N-body}

The N-body kernel computes gravitational forces between N bodies, typically applied in astrophysical simulations. The N-body kernel in BAT was created by Petrovi\v{c} et al. for use in KTT~\cite{petrovic_benchmark_2019}, as a tunable implementation of the code sample from the CUDA SDK. 
The N-body kernel follows a simple quadratic scheme where the forces between all pairs of bodies are computed every iteration. As such, the kernel is very compute intensive. 

The tunable parameters for the N-body kernel in BAT are shown in Table~\ref{tab:nbody-parameters}. 
The inner loop unroll factor parameters determine the degree to which partial loop unrolling is applied for various loops in the kernel. The \verb|outer_unroll_factor| controls the amount of work allocated to each thread. The \verb|use_soa| parameter specifies whether the input bodies are stored in an array of structures or a structure of arrays. \verb|local_mem| enables or disables the use of shared memory as a software managed cache. \verb|vector_type| is to control the number of elements loaded from memory in one instruction.
\begin{table}[ht]
    \caption{Tunable parameters -- Nbody kernel in BAT.}
    \centering
    \begin{tabular}{l|p{3cm}|l}
    \toprule
    Parameter & Values & \# \\
    \midrule
\verb|block_size| & $\{$64, 128, 256, 512$\}$ & 4 \\
\verb|outer_unroll_factor| & $\{$1, 2, 4, 8$\}$ & 4 \\
\verb|inner_unroll_factor1| & $\{$0, 1, 2, 4, 8, 16, 32$\}$ & 7 \\
\verb|inner_unroll_factor2| & $\{$0, 1, 2, 4, 8, 16, 32$\}$ & 7 \\
\verb|use_soa| & $\{$0, 1$\}$ & 2 \\
\verb|local_mem| & $\{$0, 1$\}$ & 2 \\
\verb|vector_type| & $\{$1, 2, 4$\}$ & 3 \\
 \bottomrule
    \end{tabular}
    \label{tab:nbody-parameters}
\end{table}


\subsection{Hotspot}

The Hotspot kernel included in BAT is based on the Hotspot kernel in the Rodinia Benchmark suite~\cite{che_rodinia_2009}. 
The 
kernel is part of a thermal simulation application used to estimate processor temperature based on processor architecture
and simulated power currents. The kernel iteratively solves a series of differential equations. The kernel inputs are the power and initial temperatures, the output is a grid of average temperature values spanning the chip.

To simplify the indexing scheme and increase the tunability of the kernel, we have re-implemented the Hotspot kernel in Rodinia from scratch. 
The main difference of our implementation with that of Rodinia is that our kernel can be used with any thread block dimension, can arbitrarily vary the amount of work per thread, and vary the extent to which temporal tiling is applied.

The tunable parameters for the Hotspot kernel in BAT are shown in Table~\ref{tab:hotspot-parameters}. 
\verb|block_size_x| and \verb|block_size_y| describe the thread block dimensions in x and y, the kernel uses at least 32 and at most 1024 threads.
\verb|tile_size_x| and \verb|tile_size_y| control the number of output elements computed by each thread in the x and y dimensions. \verb|temporal_tiling_factor| is the number of iterations of the stencil operation performed by a single kernel launch, for more details on the temporal tiling optimization see Hijma et al.~\cite{hijma2022}. \verb|sh_power| enables or disables the use of shared memory as a cache for storing the input power currents. \verb|blocks_per_sm| is used in the \verb|__launch__bounds()| directive in CUDA to hint the compiler to aim for a certain occupancy when running the kernel, effectively this optimization encourages the compiler to decrease register usage in the kernel.

\begin{table}[ht]
    \caption{Tunable parameters -- Hotspot kernel in BAT.}
    \centering
    \begin{tabular}{l|p{3cm}|l}
    \toprule
    Parameter & Values & \# \\
    \midrule
\verb|block_size_x| & $\{1,2,4,8,32n \mid 32n \in [32, 1024] \}$ & 37 \\
\verb|block_size_y| & $\{$1, 2, 4, 8, 16, 32$\}$ & 6 \\
\verb|tile_size_x| & $\{$n $\mid$ n $\in$ [1, 10] $\}$& 10 \\
\verb|tile_size_y| & $\{$n $\mid$ n $\in$ [1, 10] $\}$& 10 \\
\verb|temporal_tiling_factor| & $\{$n $\mid$ n $\in$ [1, 10] $\}$& 10 \\
\verb|loop_unroll_factor_t| & $\{$n $\mid$ n $\in$ [1, 10] $\}$& 10 \\
\verb|sh_power| & $\{$0, 1$\}$ & 2 \\
\verb|blocks_per_sm| & $\{$0, 1, 2, 3, 4$\}$ & 5 \\
 \bottomrule
    \end{tabular}
    \label{tab:hotspot-parameters}
\end{table}

\subsection{Pnpoly}



Pnpoly (Point-in-polygon) kernel is used by Goncalves et al.~\cite{goncalves2016spatial} as part of a geospatial database system for massive point clouds obtained through airborne LiDAR. The kernel is used to query all points within a certain outline, for example points on highways or all points within a city.
Pnpoly has been used as a benchmark kernel for autotuning in several studies~\cite{willemsen_bayesian_2021, schoonhoven2022benchmarking}. However, the Pnpoly kernel in BAT includes only the GPU kernel of the full GPU-enabled database operator. 

The tunable parameters of the Pnpoly kernel in BAT are listed in Table~\ref{tab:pnpoly-parameters}.
\verb|block_size_x| is simply the number of threads per block. \verb|tile_size| the amount of points processed by each thread. \verb|between_method| selects the algorithm to use to see if a point lies between two other points. Similarly, \verb|use_method| selects the algorithm that is used to keep track of whether the evaluated point is inside or outside of the polygon.
\begin{table}[ht]
    \caption{Tunable parameters -- Pnpoly kernel in BAT.}
    \centering
    \begin{tabular}{l|p{3cm}|l}
    \toprule
    Parameter & Values & \# \\
    \midrule
\verb|block_size_x|  & $\{ 32n | 32n \in [32, 1024] \}$ & 31 \\

 \verb|tile_size|  & $\{ 1, 2n | 2n \in [2, 20] \}$ & 11 \\
 \verb|between_method| & $\{0, 1, 2, 3\}$ & 4 \\
 \verb|use_method| & $\{0, 1, 2\}$ & 3 \\
 \bottomrule
    \end{tabular}
    \label{tab:pnpoly-parameters}
\end{table}

\subsection{Convolution}

Van Werkhoven et al.~\cite{vanWerkhoven2014optimizing}
have implemented an optimized and highly-tunable GPU-accelerated 
library for 2D Convolution operations, which has become a commonly used benchmark in autotuning~\cite{nugteren_cltune:_2015,petrovic_benchmark_2019,kerneltuner,schoonhoven2022benchmarking}.

A convolution operation computes a linear combination of weights and a range of the input 
image for each output pixel. A 2D convolution of an input image $I$ of size 
$(w\times h)$ and a convolution filter $F$ of size $(F_w\times F_h)$ computes 
an output image $O$ of size $((w-F_w)\times (h-F_h))$:
\begin{equation}\nonumber
O(x,y) = \sum\limits_{j=0}^{F_h} \sum\limits_{i=0}^{F_w} I(x+i,y+j)\times F(i,j)
\end{equation}

The tunable parameters of the Convolution kernel in BAT are listed in Table~\ref{tab:convolution-parameters}.  \verb|block_size_x| and \verb|block_size_y| describe the thread block dimensions, \verb|tile_size_x| and \verb|tile_size_y| the number of output pixels processed by each thread in the x and y dimensions.  \verb|use_padding| controls whether or not to use the padding scheme in shared memory that is used to avoid shared memory bank conflicts as described in Van Werkhoven et al.~\cite{vanWerkhoven2014optimizing}. Padding is only significant when the \verb|block_size_x| is not a multiple of number of memory banks in shared memory. Finally, \verb|read_only| controls whether or not to load input elements from global memory through read-only cache. 
\begin{table}[ht]
    \caption{Tunable parameters -- Convolution kernel in BAT.}
    \centering
    \begin{tabular}{l|p{3cm}|l}
    \toprule
    Parameter & Values & \# \\
    \midrule
 \verb|block_size_x|  & $\{$1, 2, 4, 8, 16, 32, 48, 64, 80, 96, 112, 128$\}$ & 12 \\
\verb|block_size_y|  & $\{1, 2, 4, 8, 16, 32\}$ & 6 \\
 \verb|tile_size_x|  & $\{1, 2, 3, 4, 5, 6, 7, 8\}$ & 8 \\
 \verb|tile_size_y|  & $\{1, 2, 3, 4, 5, 6, 7, 8\}$ & 8 \\
 \verb|use_padding|  & $\{0, 1\}$ & 2 \\
 \verb|read_only|  & $\{0, 1\}$ & 2 \\
 \bottomrule
    \end{tabular}
    \label{tab:convolution-parameters}
\end{table}

\subsection{Expdist}

The Expdist kernel is part of a localization microscopy applications that implements a template-free particle fusion algorithm by combining many different observations into a single super-resolution reconstruction~\cite{heydarian2018template}. The expdist kernel is used as part of the registration process where the kernel is called repeatedly to quantify the registration of two particles. The distance between two particles $t$ and $m$, given registration $M$, is computed as follows:
\begin{equation*}
D = \sum\limits_{i=1}^{K_t} \sum\limits_{j=1}^{K_m} \textrm{exp}\left( - \frac{\|\vec{x}_{t,i} - M(\vec{x}_{m,j})\|^2 }{ 2\sigma^2} \right)
\end{equation*}
The kernel operates directly on the individual localizations ($\vec{x_t}$ and $\vec{x_m}$) in each particle rather than pixelated images and takes the uncertainties in the localizations ($\sigma$) into account. The algorithm is quadratic in the number of localizations per particle and is as such very compute intensive.

The tunable parameters used in the ExpDist kernel are shown in Table~\ref{tab:expdist-parameters}.
The kernel supports two main implementations that are controlled by the \verb|use_column| parameter. When \verb|use_column| is set to 1, the kernel reduces the number of thread blocks used to perform the computation by using a fixed number of thread blocks in the y dimension, set by \verb|n_y_blocks|. \verb|use_shared_mem| use shared memory selects the way in which shared memory is used.
\begin{table}[ht]
    \caption{Tunable parameters -- ExpDist kernel in BAT.}
    \centering
    \begin{tabular}{l|p{3cm}|l}
    \toprule
    Parameter & Values & \# \\
    \midrule
\verb|block_size_x| & \{32, 64, 128, 256, 512, 1024\} & 6 \\
\verb|block_size_y| & \{1, 2, 4, 8, 16, 32\} & 6 \\
\verb|tile_size_x| & \{1, 2, 3, 4, 5, 6, 7, 8\} & 8 \\
\verb|tile_size_y| & \{1, 2, 3, 4, 5, 6, 7, 8\} & 8 \\
\verb|use_shared_mem| & \{0, 1, 2\} & 3 \\
\verb|loop_unroll_factor_x| & \{1, 2, 3, 4, 5, 6, 7, 8\} & 8 \\
\verb|loop_unroll_factor_y| & \{1, 2, 3, 4, 5, 6, 7, 8\} & 8 \\
\verb|use_column| & \{0, 1\} & 2\\
\verb|n_y_blocks| & \{1, 2, 4, 8, 16, 32, 64, 128, 256, 512, 1024\} & 11 \\
 \bottomrule
    \end{tabular}
    \label{tab:expdist-parameters}
\end{table}

\subsection{Dedispersion}

The Dedispersion kernel in BAT originates from the AMBER pipeline for the detection of single pulse astronomical transients~\cite{sclocco2020amber}.
Dedispersion is the process of reverting the dispersion of a radio signal transmitted over many frequencies through space. 
The signal component with the highest frequency $f_h$ is received at time $t_x$, while simultaneously emitted components with lower frequency arrive at $t_x + k$, where $k$ is the delay in seconds as by the dispersion equation:
\begin{equation*}
    k \approx 4150 \times DM \times \left( \frac{1}{f_i^2} \times \frac{1}{f_h^2} \right)
\end{equation*}

The kernel takes samples in time across many frequency bands (channels) as input and outputs the dedispersed samples for many different dispersion measure $DM$ values. The kernel is parallelized such that each thread can work on multiple samples and dispersion measures, while iterating over the frequency bands. As input for the BAT Dedispersion kernel, we are using the parameters from the ARTS survey on the Apertif telescope~\cite{van2022apertif}, which uses a sampling rate of 24.4 KHz, 2048 DMs, and 1536 channels.

The tunable parameters of the dedispersion kernel are shown in Table~\ref{tab:dedisp-parameters}. 
The \verb|loop_unroll_factor_channel| parameter depends on the input, as any divisor of the number of channels can be used as a partial loop unrolling factor for the inner loop in the kernel. When the loop unroll factor is 0, it is left to the CUDA compiler to decide whether or not to apply loop unrolling. \verb|tile_stride_x| controls the stride used to vary the amount of work per threads. When \verb|tile_stride_x| is 0 and \verb|tile_size_x| is larger than 1, threads will process \verb|tile_size_x| consecutive samples, when \verb|tile_stride_x| is 1 threads will process \verb|tile_size_x| samples that are each \verb|block_size_x| apart in the input. \verb|tile_stride_y| works similarly but for dispersion measures in the y-dimension.
\begin{table}[ht]
    \caption{Tunable parameters -- Dedispersion kernel in BAT.}
    \centering
    \begin{tabular}{l|p{3cm}|l}
    \toprule
    Parameter & Values & \# \\
    \midrule
\verb|block_size_y| & $\{1,2,4,8,16n \mid 16n \in [16, 512] \}$ & 36 \\
\verb|block_size_y|  & $\{4n \mid 4n \in [4,128]\}$ & 32 \\
\verb|tile_size_x|  & $\{ n | n \in [1, 16] \}$ & 16 \\
\verb|tile_size_y|  & $\{ n | n \in [1, 16] \}$ & 16 \\
\verb|tile_stride_x|  & \{0, 1\} & 2 \\
\verb|tile_stride_y|  & \{0, 1\} & 2 \\
\verb|loop_unroll_factor_channel|  & \{0, 1, 2, 3, 4, 6, 8, 12, 16, 24, 32, 48, 64, 96, 128, 192, 256, 384, 512, 768, 1536\} & 21 \\
\verb|blocks_per_sm|  & \{0, 1, 2, 3, 4\} & 5 \\
    \bottomrule
    \end{tabular}
    \label{tab:dedisp-parameters}
\end{table}

\section{Experimental Design}\label{sec:experimental-design}

\subsection{Benchmarks, Hardware and Runtime Environment}
We ran our benchmarks on four different systems with four different Nvidia GPUs.
These GPUs include the RTX 2080Ti, RTX 3060, RTX 3090 and RTX Titan. For the Pnpoly, Nbody, GEMM and Convolution benchmarks we performed an exhaustive search of the entire search space. For the Hotspot, Dedisp and Expdist benchmarks our results are based on 10 000 random configurations from the search space for each architecture. 

\section{Results and Discussion}\label{chap:results}

\subsection{Distribution of configurations}\label{subsec:distribution}

\begin{figure*}[ht]
\begin{subfigure}{.33\textwidth}
    \centering
    \includegraphics[width=\linewidth]{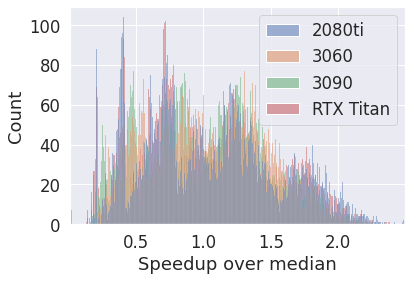}
    \caption{GEMM}
    \label{fig:my_label}
\end{subfigure}
\begin{subfigure}{.33\textwidth}
    \centering
    \includegraphics[width=\linewidth]{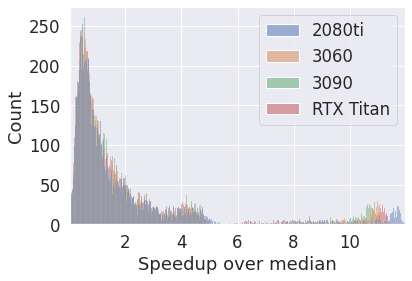}
    \caption{Hotspot}
    \label{fig:hotspot-distribution}
\end{subfigure}
\begin{subfigure}{.33\textwidth}
    \centering
    \includegraphics[width=\linewidth]{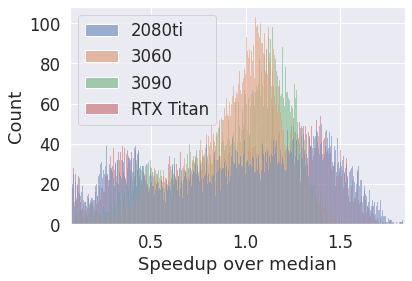}
    \caption{Dedisp}
    \label{fig:my_label}
\end{subfigure}
\\
\begin{subfigure}{.24\textwidth}
    \centering
    \includegraphics[width=\linewidth]{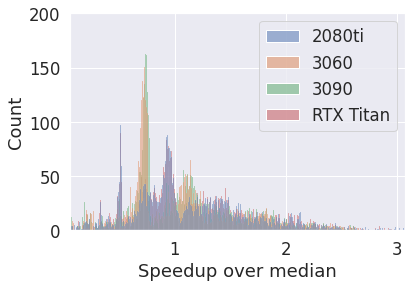}
    \caption{Convolution}
    \label{fig:my_label}
\end{subfigure}
\begin{subfigure}{.24\textwidth}
    \centering
    \includegraphics[width=\linewidth]{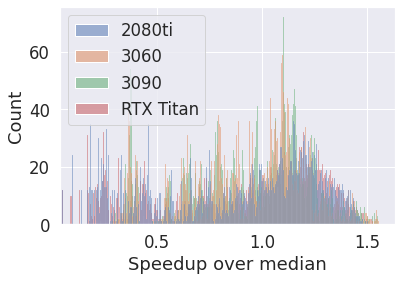}
    \caption{Pnpoly}
    \label{fig:my_label}
\end{subfigure}
\begin{subfigure}{.24\textwidth}
    \centering
    \includegraphics[width=\linewidth]{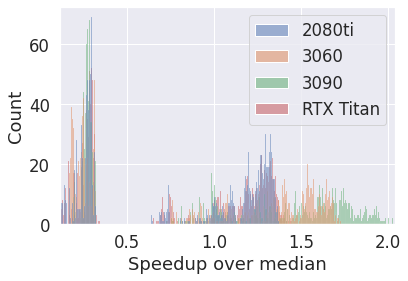}
    \caption{Nbody}
    \label{fig:nbody-distribution}
\end{subfigure}
\begin{subfigure}{.24\textwidth}
    \centering
    \includegraphics[width=\linewidth]{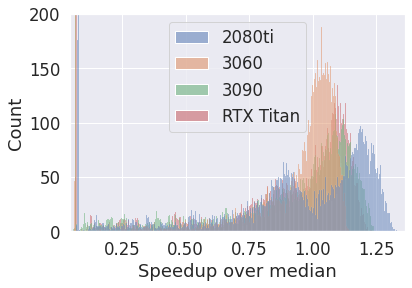}
    \caption{Expdist}
    \label{fig:my_label}
\end{subfigure}
\caption{Performance distribution of configuration for all benchmarks on all architectures}
\label{fig:distribution}
\end{figure*}

Fig~\ref{fig:distribution} shows the distribution of configurations centered around the median performing configuration. The plot extends from the worst to the best configuration. The first thing to observe is that the distribution shapes are significantly different between the different benchmarks, but similar in shape across GPUs (this agrees with results observed in other benchmark sets~\cite{olha_exploiting_2020}). Most of the benchmarks have a high density of configurations around the median and then exponential decay toward the best-performing configurations. The Hotspot benchmark in Fig.~\ref{fig:hotspot-distribution} is an outlier among the benchmarks, with a high density around the median configurations, but a cluster of very highly performing configurations giving more than 10x speedup. The Nbody distribution in Fig.~\ref{fig:nbody-distribution} also shows a distinct high-density cluster of configurations that perform very poorly.

\subsection{Convergence towards optimum using Random Search}
\begin{figure*}[ht]
\begin{subfigure}{.33\textwidth}
  \centering
  \includegraphics[width=\linewidth]{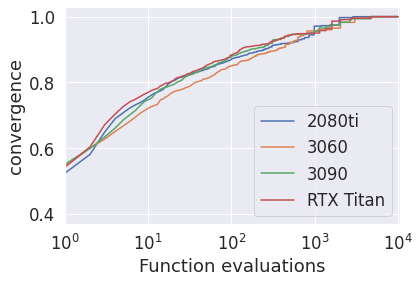}  
  \caption{GEMM}
  \label{fig:gemm-convergence}
\end{subfigure}
\begin{subfigure}{.33\textwidth}
  \centering
  \includegraphics[width=\linewidth]{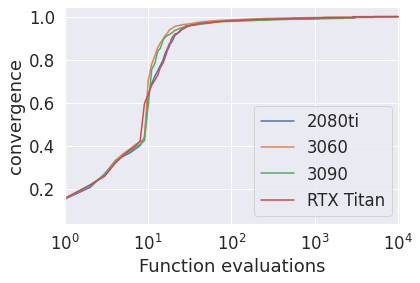}
  \caption{Hotspot}
  \label{fig:hotspot-convergence}
\end{subfigure}
\begin{subfigure}{.33\textwidth}
  \centering
  \includegraphics[width=\linewidth]{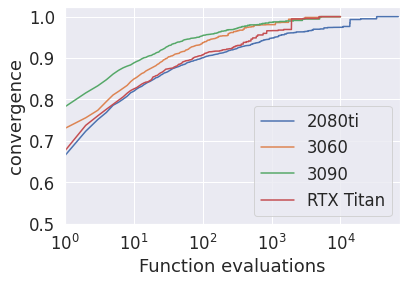}
  \caption{Dedisp}
  \label{fig:dedisp-convergence}
\end{subfigure}
\\
\begin{subfigure}{.24\textwidth}
  \centering
  \includegraphics[width=\linewidth]{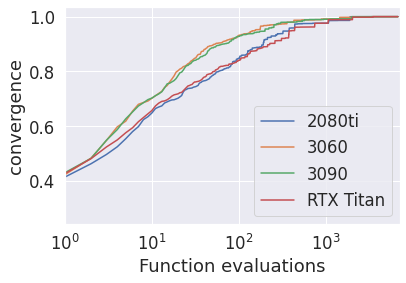}
  \caption{Convolution}
  \label{fig:convolution-convergence}
\end{subfigure}
\begin{subfigure}{.24\textwidth}
  \centering
  \includegraphics[width=\linewidth]{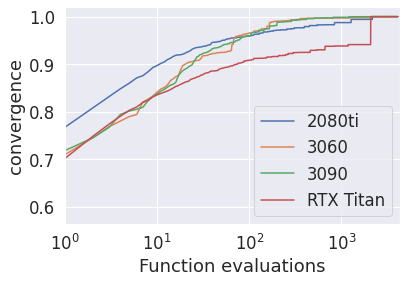}
  \caption{PnPoly}
  \label{fig:pnpoly-convergence}
\end{subfigure}
\begin{subfigure}{.24\textwidth}
  \centering
  \includegraphics[width=\linewidth]{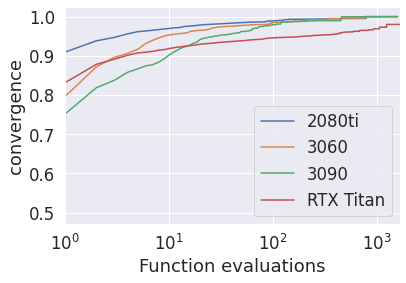}  
  \caption{Nbody}
  \label{fig:nbody-convergence}
\end{subfigure}
\begin{subfigure}{.24\textwidth}
  \centering
  \includegraphics[width=\linewidth]{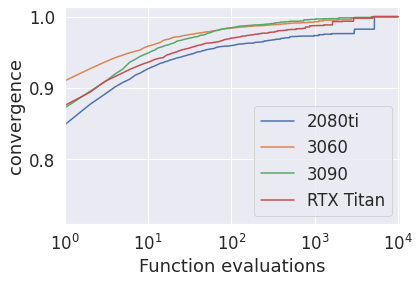}  
  \caption{Expdist}
  \label{fig:expdist-convergence}
\end{subfigure}
\caption{Convergence towards optimum for all benchmarks on all architectures}
\label{fig:convergence}
\end{figure*}

Fig.~\ref{fig:convergence} shows the convergence to the optimum configuration with relative performance (y-axis) plotted against the number of function evaluations on a Symmetric Log scale (linear scale from 0 to 1). Results are from random sampling 100 times from exhaustive or partial runs, with the median of the best evaluation plotted.


We can observe that there is a significant variance in the convergence between benchmarks, while there is less difference between GPUs. 
Again the Hotspot benchmark in Fig.~\ref{fig:hotspot-convergence} is a clear outlier, with Random Search quickly approaching a performance that is close to optimal. We stipulate that this is the due to the size of the high performing cluster shown in Section~\ref{subsec:distribution}. This cluster is likely of a sufficient size such that random search can quickly find a solution in this cluster, which is then close to optimal.

There are also significant differences in how quickly the other benchmarks converge towards the optimal. Expdist in Fig.~\ref{fig:expdist-convergence} and Nbody in Fig.~\ref{fig:nbody-convergence} achieve a 90\% optimum performance after just 10 function evaluations. For Dedisp (Fig.~\ref{fig:dedisp-convergence}) and PnPoly (Fig.~\ref{fig:pnpoly-convergence}) it takes around 100 evaluations to reach the same level. We can also see here how for the RTX Titan the PnPoly benchmark shows how a single highly performing configuration can be the source of the final jump in relative performance.

Lastly Convolution (Fig.~\ref{fig:convolution-convergence}) and GEMM (Fig.~\ref{fig:gemm-convergence}) require hundreds of configurations to exceed 90\%.

\subsection{Proportion of centrality}
\begin{figure*}[ht]
\begin{subfigure}{.33\textwidth}
    \centering
    \includegraphics[width=\linewidth]{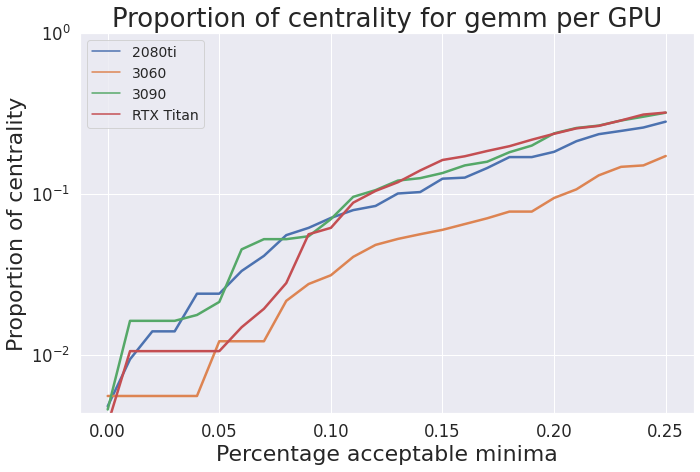}
    \caption{GEMM}
    \label{fig:my_label}
\end{subfigure}
\begin{subfigure}{.33\textwidth}
    \centering
    \includegraphics[width=\linewidth]{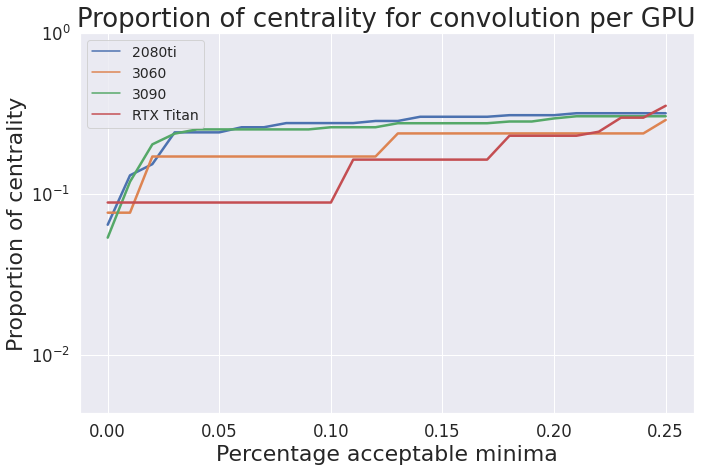}
    \caption{Convolution}
    \label{fig:my_label}
\end{subfigure}
\begin{subfigure}{.33\textwidth}
    \centering
    \includegraphics[width=\linewidth]{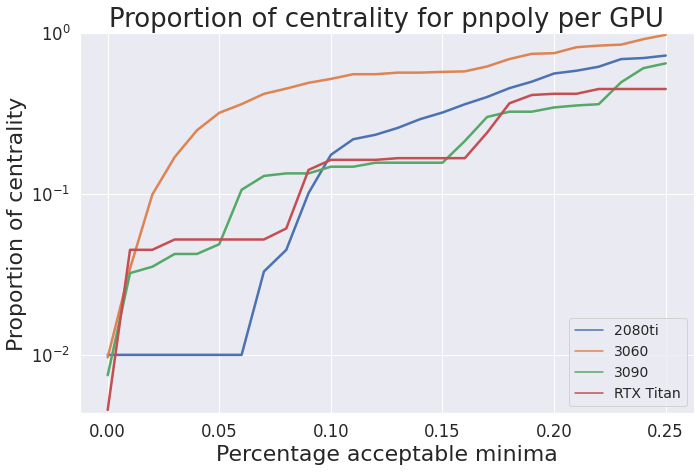}
    \caption{Pnpoly}
    \label{fig:my_label}
\end{subfigure}
\caption{Proportion of centrality for all benchmarks on all architectures}
\label{fig:centrality}
\end{figure*}

We are using the proportion of centrality metric proposed by Schoonhoven et al.~\cite{schoonhoven2022benchmarking} to calculate the search difficulty of the benchmarks.
We did not have sufficient resources to calculate the metric for the benchmarks with the largest search spaces, incl. Hotspot, Dedisp and Expdist. The results are shown in Figure~\ref{fig:centrality}. The results indicate that local search algorithms will generally find better performing configurations on the Convolution benchmark compared with GEMM and Pnpoly, which are comparatively more difficult benchmarks under this metric. This is in contrast to the results from our Random Search results, where all three benchmarks have similar trajectories towards the optimum.


\subsection{Max speedup over Median}
\begin{figure}[ht]
    \centering
    \includegraphics[width=\linewidth]{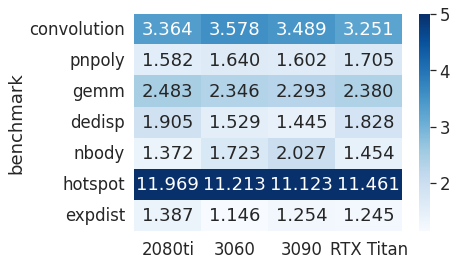}
    \caption{Max speedup over Median configuration}
    \label{fig:speedup}
\end{figure}

Fig.~\ref{fig:speedup} shows the speedup between the Median performance configuration of the search space and the best possible configuration found. While most of the benchmarks have speedups between 1.5 - 3.06x, outliers like the Hotspot benchmark have very significant speedups from 11.12 - 11.97x.

\subsection{Performance Portability}
\begin{figure*}[ht]
\begin{subfigure}{.33\textwidth}
    \centering
    \includegraphics[width=\linewidth]{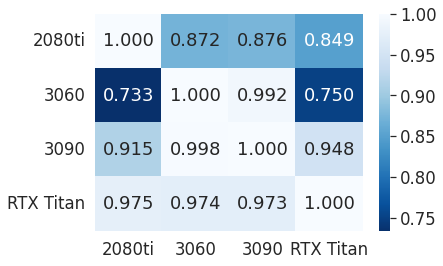}
    \caption{Convolution}
    \label{fig:portability-convolution}
\end{subfigure}
\begin{subfigure}{.33\textwidth}
    \centering
    \includegraphics[width=\linewidth]{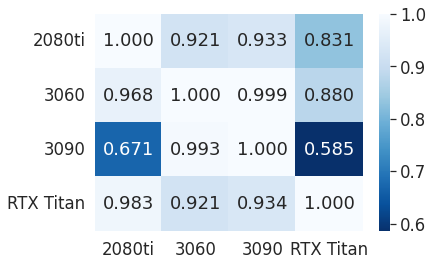}
    \caption{Pnpoly}
    \label{fig:portability-pnpoly}
\end{subfigure}
\begin{subfigure}{.33\textwidth}
    \centering
    \includegraphics[width=\linewidth]{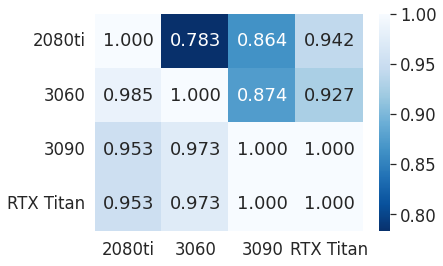}
    \caption{Nbody}
    \label{fig:my_label}
\end{subfigure}
\caption{Performance portability for exhaustively searched benchmarks on all architectures}
\label{fig:portability}
\end{figure*}

We can analyze the performance portability of configurations to measure the degree to which configurations are specifically optimized for each architecture. In Fig.~\ref{fig:portability} we show the relative performance compared with the optimal configuration for each architecture as the optimal configurations are transferred to the other architectures. The direction for this transfer is described by the optimal configuration for the GPU labeled in each row, being transferred to the different GPUs labeled on each of the columns.

In Fig.~\ref{fig:portability-pnpoly} we plot the portability for PnPoly. This shows that configurations are very portable between the RTX 3060 and RTX 3090, however configurations optimized for the RTX 3090 transfer poorly to the RTX Titan (58.5\% of optimal) and the 2080Ti (67.1\%). Similarly for the Convolution benchmark in Fig.~\ref{fig:portability-convolution} the optimal configuration for the RTX 3060 transfers poorly to the RTX 2080Ti (73.3\%) and RTX Titan (75.0\%).

\begin{figure*}[ht]
\begin{subfigure}{.33\textwidth}
    \centering
    \includegraphics[width=\linewidth]{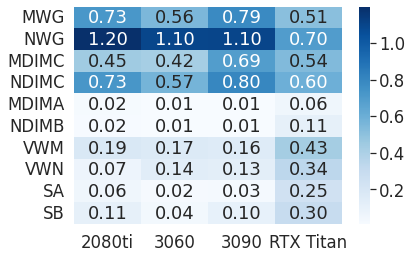}
    \caption{GEMM}
    \label{fig:features-gemm}
\end{subfigure}
\begin{subfigure}{.33\textwidth}
    \centering
    \includegraphics[width=\linewidth]{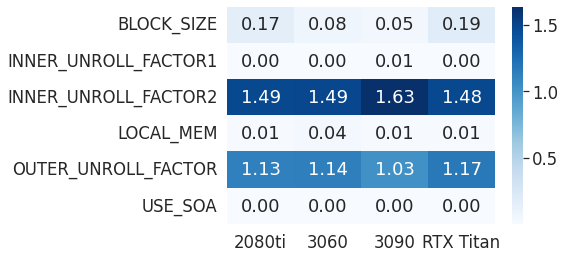}
    \caption{Nbody}
    \label{fig:features-nbody}
\end{subfigure}
\begin{subfigure}{.33\textwidth}
    \centering
    \includegraphics[width=\linewidth]{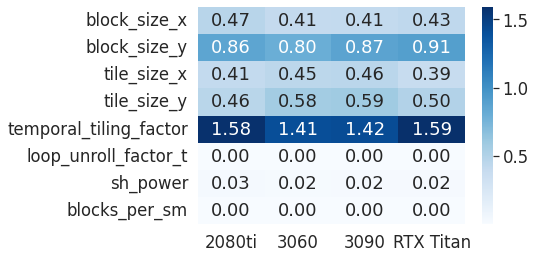}
    \caption{Hotspot}
    \label{fig:my_label}
\end{subfigure}
\\
\begin{subfigure}{.33\textwidth}
    \centering
    \includegraphics[width=\linewidth]{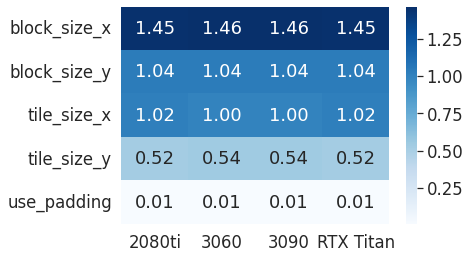}
    \caption{Convolution}
    \label{fig:my_label}
\end{subfigure}
\begin{subfigure}{.33\textwidth}
    \centering
    \includegraphics[width=\linewidth]{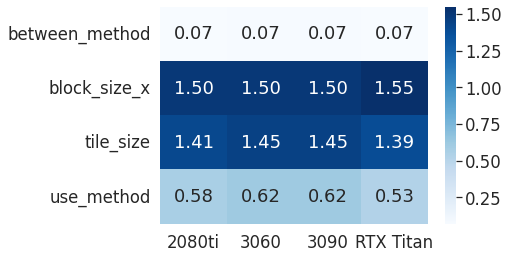}
    \caption{Pnpoly}
    \label{fig:my_label}
\end{subfigure}
\begin{subfigure}{.33\textwidth}
    \centering
    \includegraphics[width=\linewidth]{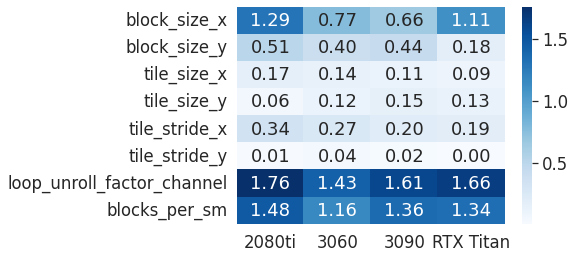}
    \caption{Dedisp}
    \label{fig:my_label}
\end{subfigure}
\\
\begin{subfigure}{.33\textwidth}
    \centering
    \includegraphics[width=\linewidth]{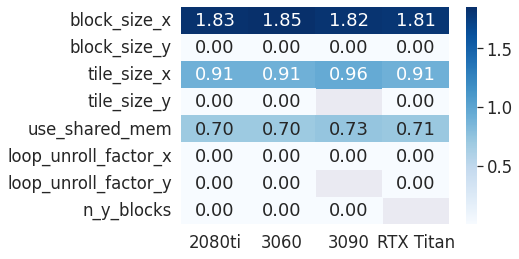}
    \caption{Expdist}
    \label{fig:my_label}
\end{subfigure}
\caption{Feature importance for all benchmarks on all architectures}
\label{fig:feature-importance}
\end{figure*}

\subsection{Feature importance}
To assess the importance of the different search parameters for the benchmark's objective, we train a Catboost Regression model over the dataset and analyze this model to investigate which features it finds useful for prediction. We use Permutation Feature Importance to then analyze the impact of each individual parameter on the model's predictive power.

Training the model on the datasets the majority of the benchmarks and GPUs that use our CatBoost model is able to predict the performance of different configurations very precisely, with an R-squared score from 0.992 and upwards for all benchmarks except Convolution, where it ranges from 0.9268 to 0.9361. 

Using this model we can then generate the feature importances in Fig.~\ref{fig:feature-importance}.
We can observe that for many of the benchmarks, especially GEMM in Fig.~\ref{fig:features-gemm} and Nbody in Fig.~\ref{fig:features-nbody}, many of the parameters do not appear to have any meaningful impact on the model's predictive performance. Although the significance of these features may not extend to GPUs that are vastly different from those tested in this study, our findings are generally consistent across the various GPUs examined.

\subsection{Feature importance impact on relevant search space}

\begin{table*}[ht]
    \caption{Search space sizes of benchmarks in BAT}
    \centering
    \begin{tabular}{c|r|r|r|r|r}
        \toprule
        Benchmark & Cardinality & Constrained & Valid & Reduced & Reduce-Constrained \\
        \midrule
        PnPoly & 4 092 & 4 092 & 3 734 - 3 774 & 4 092 & 3 734 - 3 774\\ 
        Nbody & 9 408 & 1 568 & 1 568 & 112 & 70 \\ 
        Convolution & 18 432 & 9 400 & 5 220 - 5 256 & 4 700 & 4 700\\
        GEMM & 82 944 & 17 956 & 17 956 & 17 956 & 17 956 \\
        Expdist & 9 732 096 & 540 000 & N/A & 144 & 96 \\
        Hotspot & 22 200 000 & 21 850 147 & N/A & 220 000 & 202 582 \\
        Dedisp & 123 863 040 & 107 011 905 & N/A & 3 870 720 & 3 327 135\\
        \bottomrule
    \end{tabular}
    \label{tab:space}
\end{table*}

Given our previous analysis we can reduce the search space of the benchmarks to only include those parameters that has at least 0.05 feature importance on any of the architectures. The results for the cardinality of the search spaces can be seen in Table~\ref{tab:space}. This gives an indication towards the size of the most interesting parts of the search space for the tested GPUs. Researchers using this benchmark can search over the full search space, but use this information to get better insight into how their models are able to concentrate on the most interesting parts of the search space. 



\subsection{Discussion}
The results indicate the optimization parameters in our benchmarks have a significant impact on performance. While some parameters have more impact than others, the act of optimizing some parameters interact with other parameters. We observe this behavior through the Permutation Feature Importance summing up to a value much greater than 1 for many of the benchmarks. This behavior only occurs when there are significant dependencies between features. 
Thus this provides evidence towards the need for global optimization as opposed to orthogonal search algorithms. While the importance of each parameter is generally consistent across GPU architectures, the specific values for these parameters need to be optimized for the target architecture. Our portability study shows that simply transferring the optimal configuration from one architecture to another can give as low as 58.5\% of the optimal performance, while other configurations can be ported at 99.9\% of the optimal performance. Generally this is the case for GPUs of the same family of architectures like our RTX 3060 and RTX 3090.

\section{Conclusion and Future work} \label{chap:conclusion}
HPC systems with GPU are becoming increasingly more complex and challenging to hand-tune codes for. Autotuning framworks provide means that parameterizes kernels for a range of system parameters.
BAT 2.0, the new benchmarking suite introduced in this study, provides a comprehensive framework for evaluating the performance of optimization algorithms in modern computing systems utilizing GPUs. The results of our analysis reveal that the optimization parameters have a significant impact on performance and the need for global optimization. The importance of autotuning is highlighted in the portability study, which shows that optimal performance can only be achieved by optimizing each application for a specific target architecture. The benchmarking suite facilitates the study of optimization algorithms and their effectiveness in achieving optimal performance, positioning it as a valuable tool in modern autotuning research.

A C++-based interface towards BAT is currently under development to support tuners like KTT~\cite{filipovic_autotuning_2017} and CLTune~\cite{nugteren_cltune:_2015}. 
Future work should thus include a comparison between C++-based and Python-based kernel tuners.
\section*{Acknowledgements}
The authors would like to acknowledge 
NTNU for the PhD stipend and support of our HPC-Lab that facilitated the development of this project. 
The authors would also like to thank 
Asst. Prof. Fredrik Kjolstad and his compiler group, at Stanford University, for his valuable comments and for hosting the first author during the final period of writing this article. We would also like to acknowledge PhD Candidate Tor Andre Haugdahl at NTNU
for their valuable feedback. The sixth author would also like to thank the Center for Geophysical Forecasting (SFI CGF) at NTNU and RCN (NFR proj. no. 309960) for their support during this project.


\printbibliography	

@article{sclocco2020amber,
  title={AMBER: a real-time pipeline for the detection of single pulse astronomical transients},
  author={Sclocco, Alessio and Heldens, Stijn and van Werkhoven, Ben},
  journal={SoftwareX},
  volume={12},
  pages={100549},
  year={2020},
  publisher={Elsevier}
}

@article{van2022apertif,
  title={Apertif: Phased array feeds for the Westerbork Synthesis Radio Telescope-System overview and performance characteristics},
  author={van Cappellen, WA and Oosterloo, TA and Verheijen, MAW and Adams, EAK and Adebahr, B and Braun, R and Hess, KM and Holties, H and van der Hulst, JM and Hut, B and others},
  journal={Astronomy \& Astrophysics},
  volume={658},
  pages={A146},
  year={2022},
  publisher={EDP Sciences}
}

@article{heydarian2018template,
  title={Template-free 2D particle fusion in localization microscopy},
  author={Heydarian, Hamidreza and Schueder, Florian and Strauss, Maximilian T and van Werkhoven, Ben and Fazel, Mohamadreza and Lidke, Keith A and Jungmann, Ralf and Stallinga, Sjoerd and Rieger, Bernd},
  journal={Nature methods},
  volume={15},
  number={10},
  pages={781--784},
  year={2018},
  publisher={Nature Publishing Group}
}

@article{heldens2020landscape,
  title={The landscape of exascale research: A data-driven literature analysis},
  author={Heldens, Stijn and Hijma, Pieter and Werkhoven, Ben Van and Maassen, Jason and Belloum, Adam SZ and Van Nieuwpoort, Rob V},
  journal={ACM Computing Surveys (CSUR)},
  volume={53},
  number={2},
  pages={1--43},
  year={2020},
  publisher={ACM New York, NY, USA}
}

@inproceedings{clblast,
author = {Nugteren, C.},
 title = {{CLBlast: A tuned OpenCL BLAS library}},
 booktitle = {Proceedings of the International Workshop on OpenCL},
 series = {IWOCL '18},
 year = {2018},
 pages = {5:1--5:10},
 publisher = {ACM},
}

@inproceedings{goncalves2016spatial,
  title={A spatial column-store to triangulate the Netherlands on the fly.},
  author={Goncalves, Romulo and van Tilburg, Tom and Kyzirakos, Kostis and Alvanaki, Foteini and Koutsourakis, Panagiotis and van Werkhoven, Ben and van Hage, Willem},
  booktitle={Proceedings of the 24th ACM SIGSPATIAL International Conference on Advances in Geographic Information Systems},
  pages={1--4},
  year={2016}
}

@article{vanWerkhoven2014optimizing,
  title={Optimizing convolution operations on GPUs using adaptive tiling},
author = {van Werkhoven, B. and Maassen, J. and Bal, H. E. and others},
  journal={Future Gener. Comput. Syst.},
  volume={30},
  pages={14--26},
  year={2014},
  publisher={Elsevier}
}

@article{hijma2022,
author = {Hijma, Pieter and Heldens, Stijn and Sclocco, Alessio and van Werkhoven, Ben and Bal, Henri E.},
title = {Optimization Techniques for GPU Programming},
year = {2022},
publisher = {Association for Computing Machinery},
address = {New York, NY, USA},
issn = {0360-0300},
url = {https://doi.org/10.1145/3570638},
doi = {10.1145/3570638},
journal = {ACM Comput. Surv.},
month = {nov}
}

@article{kerneltuner,
  author  = {Ben van Werkhoven},
  title   = {Kernel Tuner: A search-optimizing GPU code auto-tuner},
  journal = {Future Generation Computer Systems},
  year = {2019},
  volume  = {90},
  pages = {347-358},
  url = {https://www.sciencedirect.com/science/article/pii/S0167739X18313359},
  doi = {https://doi.org/10.1016/j.future.2018.08.004}
}

@article{schoonhoven2022benchmarking,
  title={Benchmarking optimization algorithms for auto-tuning GPU kernels},
  author={Schoonhoven, Richard and van Werkhoven, Ben and Batenburg, K Joost},
  journal={IEEE Transactions on Evolutionary Computation},
  year={2022},
  publisher={IEEE}
}

@article{olha_exploiting_2020,
  title={Exploiting historical data: Pruning autotuning spaces and estimating the number of tuning steps},
  author={Ol'ha, Jaroslav and Hozzov{\'a}, Jana and Fousek, Jan and Filipovi{\v{c}}, Ji{\v{r}}{\'\i}},
  journal={Concurrency and Computation: Practice and Experience},
  volume={32},
  number={21},
  pages={e5962},
  year={2020},
  publisher={Wiley Online Library}
}

@inproceedings{prokhorenkova_catboost_2018,
	address = {Red Hook, NY, USA},
	series = {{NIPS}'18},
	title = {{CatBoost}: unbiased boosting with categorical features},
	shorttitle = {{CatBoost}},
	urldate = {2023-01-31},
	booktitle = {Proceedings of the 32nd {International} {Conference} on {Neural} {Information} {Processing} {Systems}},
	publisher = {Curran Associates Inc.},
	author = {Prokhorenkova, Liudmila and Gusev, Gleb and Vorobev, Aleksandr and Dorogush, Anna Veronika and Gulin, Andrey},
	month = dec,
	year = {2018},
	pages = {6639--6649},
}

@article{sund_bat_2021,
	title = {{BAT}: {A} {Benchmark} suite for {AutoTuners}},
	copyright = {Copyright (c) 2021},
	issn = {1892-0721},
	shorttitle = {{BAT}},
	url = {https://ojs.bibsys.no/index.php/NIK/article/view/915},
	language = {en},
	number = {1},
	urldate = {2021-12-10},
	journal = {Norsk IKT-konferanse for forskning og utdanning},
	author = {Sund, Ingunn and Kirkhorn, Knut A. and Tørring, Jacob O. and Elster, Anne C.},
	month = nov,
	year = {2021},
	note = {Number: 1},
	pages = {44--57},
}

@inproceedings{akiba_optuna_2019,
	address = {Anchorage, AK, USA},
	series = {{KDD} '19},
	title = {Optuna: {A} {Next}-generation {Hyperparameter} {Optimization} {Framework}},
	isbn = {978-1-4503-6201-6},
	shorttitle = {Optuna},
	url = {https://doi.org/10.1145/3292500.3330701},
	doi = {10.1145/3292500.3330701},
	urldate = {2020-03-27},
	booktitle = {Proceedings of the 25th {ACM} {SIGKDD} {International} {Conference} on {Knowledge} {Discovery} \& {Data} {Mining}},
	publisher = {Association for Computing Machinery},
	author = {Akiba, Takuya and Sano, Shotaro and Yanase, Toshihiko and Ohta, Takeru and Koyama, Masanori},
	month = jul,
	year = {2019},
	pages = {2623--2631},
}

@misc{eggensperger_hpobench_2022,
	title = {{HPOBench}: {A} {Collection} of {Reproducible} {Multi}-{Fidelity} {Benchmark} {Problems} for {HPO}},
	shorttitle = {{HPOBench}},
	url = {http://arxiv.org/abs/2109.06716},
	doi = {10.48550/arXiv.2109.06716},
	urldate = {2023-01-11},
	publisher = {arXiv},
	author = {Eggensperger, Katharina and Müller, Philipp and Mallik, Neeratyoy and Feurer, Matthias and Sass, René and Klein, Aaron and Awad, Noor and Lindauer, Marius and Hutter, Frank},
	month = oct,
	year = {2022},
	note = {arXiv:2109.06716 [cs]},
}

@misc{sehic_lassobench_2022,
	title = {{LassoBench}: {A} {High}-{Dimensional} {Hyperparameter} {Optimization} {Benchmark} {Suite} for {Lasso}},
	shorttitle = {{LassoBench}},
	url = {http://arxiv.org/abs/2111.02790},
	language = {en},
	urldate = {2023-01-11},
	publisher = {arXiv},
	author = {Šehić, Kenan and Gramfort, Alexandre and Salmon, Joseph and Nardi, Luigi},
	month = jun,
	year = {2022},
	note = {arXiv:2111.02790 [cs]},
}

@misc{wang_fedhpo-b_2022,
	title = {{FedHPO}-{B}: {A} {Benchmark} {Suite} for {Federated} {Hyperparameter} {Optimization}},
	shorttitle = {{FedHPO}-{B}},
	url = {http://arxiv.org/abs/2206.03966},
	language = {en},
	urldate = {2023-01-11},
	publisher = {arXiv},
	author = {Wang, Zhen and Kuang, Weirui and Zhang, Ce and Ding, Bolin and Li, Yaliang},
	month = jun,
	year = {2022},
	note = {arXiv:2206.03966 [cs]},
}

@article{fursin_collective_2021,
	title = {Collective knowledge: organizing research projects as a database of reusable components and portable workflows with common interfaces},
	volume = {379},
	issn = {1364-503X, 1471-2962},
	shorttitle = {Collective knowledge},
	url = {https://royalsocietypublishing.org/doi/10.1098/rsta.2020.0211},
	doi = {10.1098/rsta.2020.0211},
	language = {en},
	number = {2197},
	urldate = {2023-01-11},
	journal = {Philosophical Transactions of the Royal Society A: Mathematical, Physical and Engineering Sciences},
	author = {Fursin, Grigori},
	month = may,
	year = {2021},
	pages = {rsta.2020.0211, 20200211},
}

@inproceedings{grauer-gray_auto-tuning_2012,
	address = {San Jose, CA, USA},
	title = {Auto-tuning a high-level language targeted to {GPU} codes},
	isbn = {978-1-4673-2633-9 978-1-4673-2632-2 978-1-4673-2631-5},
	url = {http://ieeexplore.ieee.org/document/6339595/},
	doi = {10.1109/InPar.2012.6339595},
	language = {en},
	urldate = {2023-01-11},
	booktitle = {2012 {Innovative} {Parallel} {Computing} ({InPar})},
	publisher = {IEEE},
	author = {Grauer-Gray, Scott and Xu, Lifan and Searles, Robert and Ayalasomayajula, Sudhee and Cavazos, John},
	month = may,
	year = {2012},
	pages = {1--10},
}

@inproceedings{zien_feature_2009,
	address = {Berlin, Heidelberg},
	series = {Lecture {Notes} in {Computer} {Science}},
	title = {The {Feature} {Importance} {Ranking} {Measure}},
	isbn = {978-3-642-04174-7},
	doi = {10.1007/978-3-642-04174-7_45},
	language = {en},
	booktitle = {Machine {Learning} and {Knowledge} {Discovery} in {Databases}},
	publisher = {Springer},
	author = {Zien, Alexander and Krämer, Nicole and Sonnenburg, Sören and Rätsch, Gunnar},
	editor = {Buntine, Wray and Grobelnik, Marko and Mladenić, Dunja and Shawe-Taylor, John},
	year = {2009},
	pages = {694--709},
}

@inproceedings{willemsen_bayesian_2021,
	title = {Bayesian {Optimization} for auto-tuning {GPU} kernels},
	doi = {10.1109/PMBS54543.2021.00017},
	booktitle = {2021 {International} {Workshop} on {Performance} {Modeling}, {Benchmarking} and {Simulation} of {High} {Performance} {Computer} {Systems} ({PMBS})},
	author = {Willemsen, Floris-Jan and van Nieuwpoort, Rob and van Werkhoven, Ben},
	month = nov,
	year = {2021},
	pages = {106--117},
}

@inproceedings{hutter_sequential_2011,
	address = {Berlin, Heidelberg},
	series = {Lecture {Notes} in {Computer} {Science}},
	title = {Sequential {Model}-{Based} {Optimization} for {General} {Algorithm} {Configuration}},
	isbn = {978-3-642-25566-3},
	doi = {10.1007/978-3-642-25566-3_40},
	language = {en},
	booktitle = {Learning and {Intelligent} {Optimization}},
	publisher = {Springer},
	author = {Hutter, Frank and Hoos, Holger H. and Leyton-Brown, Kevin},
	editor = {Coello, Carlos A. Coello},
	year = {2011},
	pages = {507--523},
}

@article{van_werkhoven_kernel_2019,
	title = {Kernel {Tuner}: {A} search-optimizing {GPU} code auto-tuner},
	volume = {90},
	issn = {0167-739X},
	shorttitle = {Kernel {Tuner}},
	url = {https://www.sciencedirect.com/science/article/pii/S0167739X18313359},
	doi = {10.1016/j.future.2018.08.004},
	language = {en},
	urldate = {2021-04-28},
	journal = {Future Generation Computer Systems},
	author = {van Werkhoven, Ben},
	month = jan,
	year = {2019},
	pages = {347--358},
}

@inproceedings{danalis_scalable_2010,
	address = {Pittsburgh, Pennsylvania, USA},
	series = {{GPGPU}-3},
	title = {The {Scalable} {Heterogeneous} {Computing} ({SHOC}) benchmark suite},
	isbn = {978-1-60558-935-0},
	url = {https://doi.org/10.1145/1735688.1735702},
	doi = {10.1145/1735688.1735702},
	urldate = {2020-05-26},
	booktitle = {Proceedings of the 3rd {Workshop} on {General}-{Purpose} {Computation} on {Graphics} {Processing} {Units}},
	publisher = {Association for Computing Machinery},
	author = {Danalis, Anthony and Marin, Gabriel and McCurdy, Collin and Meredith, Jeremy S. and Roth, Philip C. and Spafford, Kyle and Tipparaju, Vinod and Vetter, Jeffrey S.},
	month = mar,
	year = {2010},
	pages = {63--74},
}

@inproceedings{frigo_fftw_1998,
	title = {{FFTW}: an adaptive software architecture for the {FFT}},
	volume = {3},
	shorttitle = {{FFTW}},
	doi = {10.1109/ICASSP.1998.681704},
	booktitle = {Proceedings of the 1998 {IEEE} {International} {Conference} on {Acoustics}, {Speech} and {Signal} {Processing}, {ICASSP} '98 ({Cat}. {No}.{98CH36181})},
	author = {Frigo, M. and Johnson, S.G.},
	month = may,
	year = {1998},
	note = {ISSN: 1520-6149},
	pages = {1381--1384 vol.3},
}

@article{clint_whaley_automated_2001,
	series = {New {Trends} in {High} {Performance} {Computing}},
	title = {Automated empirical optimizations of software and the {ATLAS} project},
	volume = {27},
	issn = {0167-8191},
	url = {http://www.sciencedirect.com/science/article/pii/S0167819100000879},
	doi = {10.1016/S0167-8191(00)00087-9},
	language = {en},
	number = {1},
	urldate = {2020-05-22},
	journal = {Parallel Computing},
	author = {Clint Whaley, R. and Petitet, Antoine and Dongarra, Jack J.},
	month = jan,
	year = {2001},
	pages = {3--35},
}

@article{petrovic_benchmark_2019,
	title = {A {Benchmark} {Set} of {Highly}-efficient {CUDA} and {OpenCL} {Kernels} and its {Dynamic} {Autotuning} with {Kernel} {Tuning} {Toolkit}},
	url = {http://arxiv.org/abs/1910.08498},
	language = {en},
	urldate = {2020-02-27},
	journal = {arXiv:1910.08498 [cs]},
	author = {Petrovič, Filip and Střelák, David and Hozzová, Jana and Oľha, Jaroslav and Trembecký, Richard and Benkner, Siegfried and Filipovič, Jiří},
	month = oct,
	year = {2019},
	note = {arXiv: 1910.08498},
}

@inproceedings{che_rodinia_2009,
	address = {Austin, TX, USA},
	title = {Rodinia: {A} benchmark suite for heterogeneous computing},
	isbn = {978-1-4244-5156-2},
	shorttitle = {Rodinia},
	url = {http://ieeexplore.ieee.org/document/5306797/},
	doi = {10.1109/IISWC.2009.5306797},
	language = {en},
	urldate = {2020-02-26},
	booktitle = {2009 {IEEE} {International} {Symposium} on {Workload} {Characterization} ({IISWC})},
	publisher = {IEEE},
	author = {Che, Shuai and Boyer, Michael and Meng, Jiayuan and Tarjan, David and Sheaffer, Jeremy W. and Lee, Sang-Ha and Skadron, Kevin},
	month = oct,
	year = {2009},
	pages = {44--54},
}

@inproceedings{filipovic_autotuning_2017,
	address = {New York, NY, USA},
	series = {{ANDARE} '17},
	title = {Autotuning of {OpenCL} {Kernels} with {Global} {Optimizations}},
	isbn = {978-1-4503-5363-2},
	url = {http://doi.acm.org/10.1145/3152821.3152877},
	doi = {10.1145/3152821.3152877},
	urldate = {2019-09-03},
	booktitle = {Proceedings of the 1st {Workshop} on {AutotuniNg} and {aDaptivity} {AppRoaches} for {Energy} {Efficient} {HPC} {Systems}},
	publisher = {ACM},
	author = {Filipovič, Jiří and Petrovič, Filip and Benkner, Siegfried},
	year = {2017},
	note = {event-place: Portland, OR, USA},
	pages = {2:1--2:6},
}

@inproceedings{nugteren_cltune:_2015,
	address = {Turin, Italy},
	title = {{CLTune}: {A} {Generic} {Auto}-{Tuner} for {OpenCL} {Kernels}},
	isbn = {978-1-4799-8670-5},
	shorttitle = {{CLTune}},
	url = {http://ieeexplore.ieee.org/document/7328205/},
	doi = {10.1109/MCSoC.2015.10},
	language = {en},
	urldate = {2019-05-21},
	booktitle = {2015 {IEEE} 9th {International} {Symposium} on {Embedded} {Multicore}/{Many}-core {Systems}-on-{Chip}},
	publisher = {IEEE},
	author = {Nugteren, Cedric and Codreanu, Valeriu},
	month = sep,
	year = {2015},
	pages = {195--202},
}

\end{document}